\documentclass[groupedaddress,showpacs,showkeys,amssymb,eqsecnum,aps]{revtex4}     
 
\usepackage[pdftex]{graphicx}
\usepackage{amsmath} 
\usepackage{epsf}                                                                                           
\usepackage{color}                   
\usepackage{verbatim}                                                                         

\newcommand{\be}{\begin{equation}}

\newcommand{\ee}{\end{equation}}

\newcommand{\bt}{\beta}

\begin{document}                                                                                              
                                                                                                              
\title{Renormalisation of a diagonal formulation of first order Yang-Mills theory}





\author{F. T. Brandt and J. Frenkel}
\email{fbrandt@usp.br, jfrenkel@if.usp.br}
\affiliation{Instituto de F\'{\i}sica, Universidade de S\~ao Paulo, S\~ao Paulo, SP 05508-090, Brazil}
\author{D. G. C. McKeon}
\email{dgmckeo2@uwo.ca}
\affiliation{
Department of Applied Mathematics, The University of Western Ontario, London, ON N6A 5B7, Canada}
\affiliation{Department of Mathematics and Computer Science, Algoma University,
Sault St.Marie, ON P6A 2G4, Canada}

\date{\today}
                                                                                                              
\begin{abstract}
We study the BRST renormalization  of an alternative  formulation of the Yang-Mills theory, 
where the matrix-propagator of the gluon and the complementary fields is diagonal. This procedure 
involves scalings as well as non-linear mixings of the fields and sources. We show, in the Landau gauge, 
that the  BRST identities implement  a recursive proof of renormalizability to all orders. 
\end{abstract}

                                                               
\maketitle                     


\section{Introduction}
The first order formulation of gauge theories has a simple structure which contains only cubic interactions
of the fields \cite{Deser:1969wk,Okubo:1979gt}. For example, in the Yang-Mills (YM) theory, 
this formulation involves the gluon $A_\mu^a$ and the auxiliary fields $F_{\mu\nu}^a$ 
whose dynamics is described by the Lagrangian 
\be\label{eq1.1}
         {\cal L}^{1}_{YM} = \frac 1 4 F_{\mu\nu}^{ a} F^{a\, \mu\nu}
-\frac 1 2 F^{a\,\mu\nu} \left(\partial_\mu A^{a}_{\nu}  - \partial_\nu A^{a}_{\mu}
         + g f^{abc} A^{b}_\mu A^{c}_\nu \right) ,
\ee
where $f^{abc}$ are the SU(N) structure constants and $g$ is the coupling constant.
This form has a single vertex $\langle FAA \rangle$ but  leads to a rather complex  non-diagonal matrix-propagator, 
containing the 
$\langle AA\rangle$, 
$\langle FF\rangle$, and the mixed $\langle FA\rangle$ and $\langle AF\rangle$
propagators \cite{Brandt:2015nxa}.
At the classical level, \eqref{eq1.1} is equivalent to the usual second order Lagrangian 
\be\label{eq1.2}
{\cal L}^{2}_{YM} = -\frac 1 4 \left(\partial_\mu A^{a}_\nu  - \partial_\nu A^{a}_\mu  + g f^{abc} A^{ b}_\mu A^{ c}_\nu \right)^2
\ee
This may be seen by using the equation of motion for $F_{\mu\nu}^a$ and substituting it back in Eq. \eqref{eq1.1}.

The issue of  the equivalence of these formulations at the quantum level and of the  renormalizability of the
first order formulation of YM theory has been previously studied from various points
of view  \cite{McKeon:1994ds,Martellini:1997mu,Andrasi:2007dk,costello:2011b, Frenkel:2017xvm,Batalin:2018enf}.

On the other hand, it has been noticed that \cite{Brandt:2016eaj}, if we make in \eqref{eq1.1} the shift
\be\label{eq1.3}
F_{\mu\nu}^a = H_{\mu\nu}^a + \partial_\mu A^{ a}_\nu  - \partial_\nu A^{ a}_\mu
\ee
one obtains the Lagrangian
\be\label{eq1.4}
{\cal L}^{I}_{YM} = \frac 1 4 H_{\mu\nu}^a  H^{a\, \mu\nu} - \frac 1 4 
\left(\partial_\mu A^{a}_\nu  - \partial_\nu A^{a}_\mu\right)^2   -\frac 1 2 g f^{abc} 
\left(H_{\mu\nu}^a + \partial_\mu A_\nu^a  - \partial_\nu A_\mu^a  \right) A^{b\,\mu} A^{c\, \nu} 
\ee
which involves two cubic vertices $\langle AAA\rangle $ and $\langle HAA\rangle$, as well as two simple propagators
$\langle AA \rangle$ and $\langle HH\rangle$. 
This form
leads to a diagonal matrix-propagator of the $A$ and $H$ fields,
which is an useful feature in perturbation theory.   
Thus, the formulation \eqref{eq1.4} provides a convenient basis of independent fields, which is not manifest 
in the mixed first order form \eqref{eq1.1}.

The purpose of this work  is to study the renormalizability of this formulation of YM theory. 
We employ the  Landau gauge which leads to significant simplifications, due to the transversality 
of the gluon propagator. In section 2 we give the one-loop results for the ultraviolet divergent 
Green functions. In section 3 we  introduce the BRST identities which preserve the gauge invariance of 
the theory and derive the counter-terms necessary to renormalise it to one-loop order. The one-loop 
renormalization is performed in section 4, by requiring that the counter-terms cancel the divergences 
which arise in perturbation theory.
This leads to the expected result for the $\beta$-function, which accounts for asymptotic freedom.
Using the BRST identities, we give in section 5 a proof of the renormalizability 
to all orders. 
The proof is based on the fact that  
the bare and the renormalized fields and sources are related by non-linear mixings, as well as
by scalings (see Eqs. \eqref{eqs5.3}). We finally obtain, in the Landau gauge, the complete renormalized Lagrangian \eqref{eq5.7}.

\section{The ultraviolet divergences}  

The basic Lagrangian of the theory, in covariant gauges, can be written as 
\be\label{eq2.1}
{\cal L}^\prime  =  {\cal L}_0 - \frac{1}{2\xi}   (\partial^\mu A_\mu)^2,
\ee
where $\xi$ is the gauge fixing parameter which vanishes in the Landau gauge
(we use a color vector notation, with   $A\cdot B = A^a B^a$
and $(A\wedge B)^a = f^{abc} A^b B^c$, leaving the color indices $a$, $b$, $\dots$ understood) and
\be\label{eq2.2}
{\cal L}_0  =  {\cal L}^I_{YM} + \left(\partial^\mu\bar\eta + U^\mu\right)\cdot (D_\mu \eta) 
- \frac g 2 V\cdot(\eta\wedge\eta) + 
g \Omega^{\mu\nu}\cdot\left[(H_{\mu\nu} +A_\mu\partial_\nu -A_\nu \partial_\mu)\wedge\eta\right].    
\ee
Here, ${\cal L}^I_{YM}$ is given by \eqref{eq1.4}, $D_\mu$ is the covariant derivative and $\eta$, $\bar\eta$ denote 
the anticommuting ghost fields. The sources $U^\mu$, $V$ and $\Omega^{\mu\nu}$ 
are introduced for the purpose of setting up the BRST
Identities. The coefficients of these sources are invariant under the BRST transformations    
\be\label{eq2.3}
\delta H_{\mu\nu} = g(H_{\mu\nu}  + A_\mu\partial_\nu  - A_\nu \partial_\mu)\wedge\eta  \tau;
\;\delta A_\mu = (D_\mu \eta)\tau;\; \delta\eta = -\frac g 2 (\eta\wedge\eta)\tau
\ee
which can be verified using the Jacobi identity ($\tau$ is an infinitesimal Grassmann quantity).
The Feynman rules obtained from  \eqref{eq2.1} are given in Appendix A.
The divergent contributions from the one-loop Feynman graphs in a general co-variant gauge are summarized
in Appendix B.
Here, we give the results obtained in the Landau gauge for the one-loop divergencies.
Using dimensional regularisation in $4-2\epsilon$ dimensions, we can express these contributions in terms 
of the constant
\be\label{eq2.4}
d_N = \frac{g^2 N}{16\pi^2\epsilon},
\ee
which is divergent in four space-time dimensions.

The divergent parts of the $A$ and $H$ self-energies and of the mixed ($A H$) graphs
may be written in momentum space as
\be\label{eq2.5}
\Pi_{\mu\nu}^{ab}(k) = -\frac{13}{6} d_N \, \left[ i {\delta^{ab}} (k_\mu k_\nu - k^2 \eta_{\mu\nu})\right],  
\ee
\be\label{eq2.6}
\Pi_{\mu\nu;\;\alpha\beta}^{ab} = \frac{1}{2} d_N 
\left[\frac{i \delta^{ab}}{4}{(\eta_{\mu\alpha}\eta_{\nu\beta} - \eta_{\mu\beta}\eta_{\nu\alpha} )}\right] , 
\ee
\be\label{eq2.7}
\Pi_{\gamma;\, \mu\nu}^{ab}(k)=  \frac{5}{8}  d_N \, \left[ \delta^{ab} (\eta_{\gamma\mu}  k_\nu - \eta_{\gamma\nu} k_\mu)\right]. 
\ee
The tensors inside the square brackets in Eq. \eqref{eq2.5} and \eqref{eq2.6} are identical to the momentum space representation of 
the quadratic terms in $(i S)$. The square bracket in Eq. \eqref{eq2.7} is the same as the momentum space representation of the 
mixed bi-linear term when we employ the Lagrangian in Eq. \eqref{eq1.1}. 
The $(AA)$ and $(AH)$ contributions are transverse, which is a consequence of the BRST identities. 

The divergent parts of the three-point $\langle AAA\rangle$  and  $\langle HAA\rangle$  vertices have the forms
(see the Appendix A for the explicit forms of the tree level vertices)
\be\label{eq2.8}
V^{abc}_{\mu\nu\gamma} = - \frac{d_N}{6}  \,
\left({{{}^c{}^{,\alpha,} {}^{\,k_3}}  \atop {{{}_a{}_{,\mu,}{}_{\,k_1} \;}
\includegraphics[scale=0.6,trim=0.45 0.2cm 0 0]{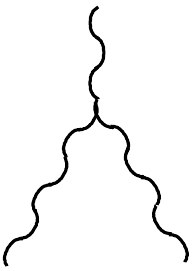}{\;{}_b{}_{,\nu}} {}_{\,k_2}}  } \right)
\ee
\be\label{eq2.9}
V^{ab\;c}_{\alpha,\beta;\mu\nu} = - \frac{d_N}{4} \,
\left( {{{}^c{}^{,\mu\nu} }  \atop {{{}_a{}_{,\alpha} \;}\includegraphics[scale=0.6,trim=0.45 0.2cm 0 0]{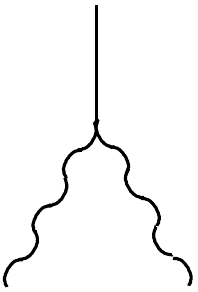}{\;{}_b{}_{,\beta} }}}  \right)
\ee
The divergent part of the one-loop four-gluon vertex $\langle AAAA\rangle$   has the structure
\be\label{eq2.10}
V^{abcd}_{\lambda\mu\nu\rho} = c_{AAAA} d_N
\left(
{                    
{{}^d{}^{,\rho} \;\;\;\;\;\;\;\;\;\;\;\;\;\;\;\,
{}^c{}^{,\nu} 
 }  \atop {{{}_a{}_{,\lambda} \;}\includegraphics[scale=0.6,trim=0.45 0.2cm 0 0.5cm]{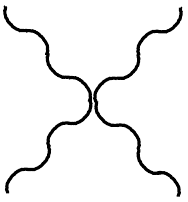}{\;{}_b{}_{,\mu} }}}
\right),
\ee
with $c_{AAAA} = \frac 1 3 $. In the Appendix B  we present some details of the calculation of
this divergent term, including the gauge parameter dependence. 

The ultraviolet behaviour of the ghost sector is similar to that in the usual second-order theory.
Thus, the divergent part of the ghost self-energy and of the $U$-$\eta$ mixing  is the same  
\be \label{eq2.11}
-\frac 3 4 d_N (\partial^\mu \eta+U^\mu)(\partial_\mu\eta) 
\ee
while the divergent parts of the  vertices involving the sources $U^\mu$, $V$, the gluon 
and ghosts vanish in the  Landau gauge.

Several divergent vertices involving the source  $\Omega^{\mu\nu}$,
the fields $A^{\mu}$   or  $H^{\mu\nu}$  and the ghosts vanish in this gauge.
The only divergent vertex involving the sources has the structure 
\be\label{eq2.12}
c g d_N \Omega^{\mu\nu} \cdot(A_\mu \partial_\nu - A_\nu\partial_\mu)\wedge \eta
\ee
where $c = -1/4$ to one-loop order.

\section{BRST identities and the counter-terms}  
The basic action is (see Eq. \eqref{eq2.2})
\be\label{eq3.1}
\Gamma_0 = \int d^4 x {\cal L}_0(x).
\ee
Let $\Gamma$ be the complete effective action which generates the one-particle irreducible Green's functions.
The BRST identities can be obtained in a similar manner to that used in the usual
second order theory, by starting from the generating functional for connected Green's functions
and making a Legendre transformation \cite{brs74,KlubergStern:1974xv}.
In this way, one gets the BRST equation
\be\label{eq3.2}
\Gamma\star\Gamma \equiv \int d^4 x\left[
\frac{\delta\Gamma}{\delta H_{\mu\nu}}\cdot\frac{\delta\Gamma}{\delta \Omega^{\mu\nu}}+
\frac{\delta\Gamma}{\delta A_{\mu}}\cdot\frac{\delta\Gamma}{\delta U^{\mu}}+
\frac{\delta\Gamma}{\delta \eta}\cdot\frac{\delta\Gamma}{\delta V}\right] = 0.
\ee
This identity reflects the gauge invariance for the theory. For example, using \eqref{eq2.3} 
and  \eqref{eq3.1},  \eqref{eq3.2} leads to zeroth order to the relation
\be\label{eq3.3}
\int d^4 x\left[
\frac{\delta\Gamma_0}{\delta H_{\mu\nu}} \cdot\delta H_{\mu\nu} +
\frac{\delta\Gamma_0}{\delta A_{\mu}} \cdot\delta A_{\mu} + 
\frac{\delta\Gamma_0}{\delta \eta} \cdot\delta \eta
\right] = 0
\ee
which implies that $\Gamma_0$ is gauge invariant.

If the theory is renormalisable, one could choose the counter-terms so as to cancel the 
UV divergences to one-loop order:  $\Gamma_1^C = -\Gamma_1^{div}$. 
Then, it  would follow from \eqref{eq3.2} that 
\be\label{eq3.4}
\Gamma_0\star\Gamma_1^C + \Gamma_1^C \star\Gamma_0 \equiv \Delta\Gamma_1^C = 0
\ee
where
\be\label{eq3.5}
\Delta \equiv \int d^4 x\left[
\frac{\delta\Gamma_0}{\delta H_{\mu\nu}}\cdot\frac{\delta }{\delta \Omega^{\mu\nu}}+
\frac{\delta\Gamma_0}{\delta \Omega_{\mu\nu}}\cdot\frac{\delta }{\delta H^{\mu\nu}}+
\frac{\delta\Gamma_0}{\delta A_{\mu}}\cdot\frac{\delta }{\delta U^{\mu}}+
\frac{\delta\Gamma_0}{\delta U_{\mu}}\cdot\frac{\delta }{\delta A^{\mu}}+
\frac{\delta\Gamma_0}{\delta \eta}\cdot\frac{\delta }{\delta V}+
\frac{\delta\Gamma_0}{\delta V}\cdot\frac{\delta }{\delta \eta}
\right].
\ee
It can be verified that  $\Delta$ is  nilpotent:  $\Delta^2 = 0$. Therefore, a class of solutions
of the equation (3.4) may have the form
\be\label{eq3.6}
\Gamma_1^{C_1} = \Delta G
\ee
where $G$ is a polynomial in the fields and sources, which is a Lorentz scalar and invariant under rigid colour transformations.
A general $G$ with these properties, which has the
correct ghost number $1$ and mass dimensions $-1$ can be written as 
\begin{eqnarray}\label{eq3.7}
G & = & \int d^4 x\left[\left(Z_A^{1/2}-1\right)A_\mu\cdot U^\mu + 
\left(Z_H^{1/2}-1\right)H_{\mu\nu}\cdot \Omega^{\mu\nu} + 
Z_{AH}\left(\partial_\mu A_\nu-\partial_\nu A_\mu\right)\cdot\Omega^{\mu\nu} \right. \nonumber \\ 
& + & \left. g Z_{AAH} \left(A_\mu\wedge A_\nu  \right)\cdot \Omega^{\mu\nu}  
+ g \left(1- Z_\eta Z_{A}^{1/2}\right) \eta \cdot V 
 \right].
\end{eqnarray}
We have omitted here a structure of the form $\eta\cdot\left(\Omega^{\mu\nu}\wedge \Omega_{\mu\nu}\right)$ since, in the Landau gauge,
this would not generate divergent contributions to $\Gamma_1^{C_1}$.
The coefficients $Z_A$,  $Z_H$,   and $Z_\eta$,
which are of order  $1+{\cal O}(\hbar)$, will generate scaling
while the coefficients $Z_{AH}$ and $Z_{AAH}$ 
which are of order ${\cal O}(\hbar)$, will generate mixing.

A second type of counter-terms may be obtained by differentiating \eqref{eq3.1} with respect
to the coupling constant $g$
\be\label{eq3.8}
\Gamma_1^{C_2} = g Z^\prime_g \frac{d\Gamma_0}{d g},
\ee
where $Z^\prime_g$  denotes a rescaling of the coupling constant.

A third type of solutions of \eqref{eq3.4} consists of terms which are explicitly gauge invariant
\begin{eqnarray}\label{eq3.9}
\Gamma_1^{C_3} & = & \int d^4 x\left[ z  \left( H_{\mu\nu} + \partial_\mu A_\nu  - \partial_\nu A_\mu \right)^2
+{z^\prime}\left(\partial_\mu A_\nu - \partial_\nu A_\mu + g A_\mu \wedge A_\nu \right)^2 \right. \nonumber \\ 
&+& \left.  {z^{\prime\prime}}\left( H^{\mu\nu} + \partial^\mu A^\nu  - \partial^\nu A^\mu \right)\cdot
\left(\partial_\mu A_\nu - \partial_\nu A_\mu + g A_\mu \wedge A_\nu \right)
\right] .
\end{eqnarray} 
However, it turns out that the coefficients  $z$, $z^\prime$ and $z^{\prime\prime}$ are 
redundant since these can be absorbed into the other counter-terms \cite{Frenkel:2017xvm}. From now on we will,
for simplicity, put $z=z^\prime=z^{\prime\prime}=0$. 
Adding   $\Gamma_0$      to the sum of above types of contributions, we get 
\be\label{eq3.10}
\Gamma_0 + \Gamma_1^C = \Gamma_0 + \Gamma_1^{C_1} + \Gamma_1^{C_2}  = \int d^4 x\left[
{\cal L}^i(x)  + {\cal L}^{ii}(x)
\right]  ,
\ee
${\cal L}^i(x)$  and  ${\cal L}^{ii}(x)$   being, respectively, the source-free and the source-dependent parts.
After a straightforward calculations, we obtain to one-loop order
\begin{eqnarray}\label{eq3.11}
{\cal L}^i &=& \frac 1 4 Z_H H^{\mu\nu} \cdot H_{\mu\nu}
+ \frac 1 2 Z_{AH} H^{\mu\nu} \cdot \left(\partial_\mu A_\nu  - \partial_\nu A_\mu\right) \nonumber \\
&-& \frac 1 4 Z_{A} \left(\partial_\mu A_\nu  - \partial_\nu A_\mu\right)^2 
-\frac{g}{2}\left( Z^\prime_g Z_A Z_H^{1/2} - Z_{AAH} \right)H^{\mu\nu}\cdot \left(A_\mu\wedge A_\nu\right) 
\nonumber \\
&-&\frac{g}{2}\left( Z^\prime_g Z_A^{3/2} + Z_{AH} \right) 
\left(\partial_\mu A_\nu  - \partial_\nu A_\mu\right) \cdot \left(A^\mu\wedge A^\nu\right) 
\nonumber \\
&-&\frac{g^2}{2} Z_{AAH} \left(A_\mu\wedge A_\nu\right)\cdot \left(A^\mu\wedge A^\nu\right) 
+ \partial^\mu \bar\eta \cdot \left( Z_\eta \partial_\mu + g A_\mu \wedge \right) \eta 
\end{eqnarray}
 and
\begin{eqnarray}\label{eq3.12}
{\cal L}^{ii} &=&  g  \Omega^{\mu\nu} \cdot \left(H_{\mu\nu} + A_\mu \partial_\nu  - A_\nu \partial_\mu\right)\wedge\eta 
- \frac g 2 V\cdot\left(\eta\wedge\eta\right)+U^\mu\cdot\left(Z_\eta \partial_\mu + g A_\mu\wedge\right) \eta
\nonumber \\
&+& g \left(Z_{AH} + Z_A^{1/2} - Z_H^{1/2}-Z_{AAH}\right)\Omega^{\mu\nu} \cdot 
\left(A_\mu \partial_\nu  - A_\nu \partial_\mu A_\mu\right)\wedge\eta ,
\end{eqnarray}
where we used the fact (see below) that, in the Landau gauge
\be\label{eq3.13}
Z_\eta Z^\prime_g Z_A^{1/2} = \tilde Z_3 Z_g Z_3^{1/2} = 1
\ee
$\tilde Z_3$, $Z_3$ and  $Z_g$ 
being, respectively, the ghost, gluon and coupling renormalization constants in the standard YM theory.           
In the next section, it will be shown that the Lagrangians \eqref{eq3.11} and \eqref{eq3.12} may 
also be obtained by making in \eqref{eq2.2} appropriate rescalings and mixings of the fields and sources.

\section{One-loop renormalization and the   $\beta$      function}
The renormalization is performed by requiring that the counter-term cancel the divergences
arising in the evaluation of the Feynman diagrams. To this end we note that, since the ghost
self-energy and the ghost-gluon vertex are the same as in the conventional second order
YM theory, we obtain the relations
\begin{subequations}
\begin{eqnarray} 
Z_\eta &=& \tilde Z_3 = 1+ \frac{3d_N}{4} \label{eq4.1}\\
Z^\prime_g Z_A^{1/2} &=&  Z_g Z_3^{1/2} \label{eq4.2}
\end{eqnarray}
\end{subequations}
Using these relations, we see that the last counter-term in \eqref{eq3.11} cancels the divergent source-free ghost 
terms in \eqref{eq2.11}. Requiring that the other counter-terms in \eqref{eq3.11} to cancel 
the one-loop divergences coming from the source-free graphs, and using the results given in section 2, yields the relations
\begin{subequations}
\begin{eqnarray} 
Z^\prime_g &=& 1-\frac{11}{6} d_N = Z_g \label{eq4.3}\\
Z_A &=& 1+\frac{13}{6} d_N = Z_3 \label{eq4.4}\\
Z_H &=& 1-\frac{1}{2} d_N  \label{eq4.5}\\
Z_{AH} &=& - \frac{5d_N}{4}\label{eq4.6}\\
Z_{AAH} &=& -\frac{d_N}{6}  = -\frac 1 2 C_{AAAA}\label{eq4.7}
\end{eqnarray}
\end{subequations}
Eq. \eqref{eq4.3} leads to the correct result for the  $\beta$-function, which is responsible for the asymptotic freedom. 

We note here that the last condition in \eqref{eq4.7} is a consequence of the  BRST identities and of the requirement 
that the counter-terms cancel the divergences in the Feynman diagrams.
We have verified this relation by explicit calculation to one-loop order (see \eqref{eq2.10}).

We now consider the renormalization of the one-loop divergences coming from the graphs
involving the sources.  The counter-term   $Z_\eta$  in \eqref{eq3.12} cancels, by \eqref{eq4.1}, 
the corresponding divergent contribution in \eqref{eq2.11}.
Finally, a comparison between the last term in \eqref{eq3.12} and \eqref{eq2.12} leads to the relation
\be\label{eq4.8}
Z_{AH}+Z^{1/2}_A - Z_H^{1/2} - Z_{AAH} = \frac{d_N}{4}.
\ee
We have verified explicitly that this condition  is satisfied in the Landau gauge. 

The counter-terms in \eqref{eq3.11} and  \eqref{eq3.12} may also be obtained from 
the tree Lagrangian \eqref{eq2.2} by substituting here the renormalized fields and sources by the bare ones.
These are found by rescaling and mixing the fields and sources in \eqref{eq2.2} in a way consistent with BRST  identities.
The required transformations may be written in the Landau gauge  as 
\begin{subequations}
\begin{eqnarray}
g^B &=&  Z^\prime_g g;\; A_\mu^B = Z_A^{1/2} A_\mu;\; (\eta^B,\bar\eta^B)  = Z_\eta^{1/2} (\eta,\bar\eta)\label{eq4.9}\\
H_{\mu\nu}^B &=&  Z_H^{1/2} H_{\mu\nu}  + Z_{AH}\left(\partial_\mu A_\nu-\partial_\nu A_\mu\right) 
+ g Z_{AAH} A_\mu\wedge A_\nu \label{eq4.10}\\
U_{\mu}^B &=&  Z_\eta^{1/2} U_{\mu}  - 2 Z_{AH} \partial^\nu \Omega_{\mu\nu} 
- 2 g Z_{AAH} A^\nu\wedge \Omega_{\mu\nu} \label{eq4.11}\\
V^B & = & Z_A^{1/2} V;\;\; \Omega^B_{\mu\nu} = \left( Z_\eta^{1/2}+Z_A^{1/2} - Z_H^{1/2}   \right)\Omega_{\mu\nu} \label{eq4.12}
\end{eqnarray}
\end{subequations}
Using the above transformations, we get to one-loop order  that 
\be\label{eq4.13}
\Gamma^R_{(1)} \equiv \left( \Gamma_0+\Gamma^C_1 \right)\left(g;A,\eta,\bar\eta,H;U,V,\Omega\right) 
=\Gamma_0\left(g^B;A^B,\eta^B,\bar\eta^B,H^B;U^B,V^B,\Omega^B\right) 
\ee
where $\Gamma_0+\Gamma_1^C$   is given by \eqref{eq3.10}--\eqref{eq3.12} and  $\Gamma_{(1)}^R$
represents the renormalized action which includes terms of zeroth and first order in $\hbar$.      

\newpage

\section{Renormalization to higher orders}
The above procedure for renormalising this version of the Yang-Mills theory to one-loop order
may be generalised, recursively, to higher orders in perturbation theory. We must show that we can 
rescale/mix the fields and the sources so that the renormalized action leads to finite Green functions and 
BRST identities are preserved at each order. The proof that this holds
to all orders may be made by induction, employing a rather similar argument to the one used 
in references  \cite{taylor:1976b,itzykson:1980b}. In order for the recursive procedure to  work,
one assumes that the  renormalized action,  $\Gamma^R_{(n)} = \Gamma_0+\Gamma_n^C$ 
which includes terms up to the order $\hbar^n$,   satisfies 
\be\label{eq5.1}
\Gamma^R_{(n)} \star \Gamma_{(n)}^R = 0 
\ee
and  one must prove that  $\Gamma^R_{(n+1)}$ obeys a similar BRST equation. 
This may be shown using a theorem which generalizes \eqref{eq4.13} to all orders.
It states that the complete renormalized action
\be\label{eq5.2}
\Gamma^R \equiv \left( \Gamma_0+\Gamma^C \right)\left(g;A,\eta,\bar\eta,H;U,V,\Omega\right) 
=\Gamma_0\left(g^B;A^B,\eta^B,\bar\eta^B,H^B;U^B,V^B,\Omega^B\right) 
\ee
satisfies the BRST equation, provided that the bare quantities are related to the renormalized ones by the 
rescaling and mixing equations
\begin{subequations}\label{eqs5.3}
\begin{eqnarray}
g^B &=&  Z^\prime_g g;\; A_\mu^B = Z_A^{1/2} A_\mu;\; (\eta^B,\bar\eta^B)  = Z_\eta^{1/2} (\eta,\bar\eta)\label{eq5.3}\\
H_{\mu\nu}^B &=&  Z_H^{1/2} H_{\mu\nu}  + Z_{AH}\left(\partial_\mu A_\nu-\partial_\nu A_\mu\right) 
+ g Z_{AAH} A_\mu\wedge A_\nu \label{eq5.4}\\
U_{\mu}^B &=&  Z_\eta^{1/2} U_{\mu}  - 2 \left(\frac{Z_{\eta}}{Z_H}\right)^{1/2} 
\left(Z_{AH} \partial^\nu \Omega_{\mu\nu}  + g Z_{AAH} A^\nu\wedge \Omega_{\mu\nu}\right) \label{eq5.5}\\
V^B & = & Z_A^{1/2} V;\;\; \Omega^B_{\mu\nu} = \left( Z_\eta Z_A /Z_H \right)^{1/2}\Omega_{\mu\nu} \label{eq5.6}
\end{eqnarray}
\end{subequations}
These transformations reduce at one loop order to those in Eqs. \eqref{eq4.9}--\eqref{eq4.12}.
The proof of the above theorem in the Landau gauge is given in Appendix C.
When the relations \eqref{eq5.3}-\eqref{eq5.6} are substituted in \eqref{eq5.2}, we get the all orders 
renormalized action. The general renormalized Lagrangian in the Landau gauge may be written as
\begin{eqnarray}\label{eq5.7}
{\cal L} ^R &=& \frac{1}{4} H^{B\,\mu\nu} \cdot H^B_{\mu\nu} - \frac{Z_A}{4}\left(\partial_\mu A_\nu-\partial_\nu A_\mu\right)^2
-g \frac{Z^\prime_gZ_A}{2}\left[H^B_{\mu\nu} + Z_A^{1/2} \left(\partial_\mu A_\nu-\partial_\nu A_\mu\right)\right]\wedge 
\left(A^\mu\wedge A^\nu\right)  \nonumber \\
&+&\left( \partial^\mu\bar\eta + U^\mu \right)\cdot \left( Z_\eta \partial_\mu + A_\mu\wedge\right)\eta
-\frac g 2 V\cdot \left(\eta\wedge\eta\right) + g \Omega^{\mu\nu}\cdot H_{\mu\nu}\wedge\eta \nonumber \\
&+&   g\left[ 1+ Z^\prime_g Z_A^{1/2}\left( Z_{AH} + Z_A^{1/2} - Z_H^{1/2}\right) -Z_{AAH}\right]\Omega^{\mu\nu}
\cdot \left(A_\mu\partial_\nu - A_\nu\partial_\mu\right)\wedge\eta
\end{eqnarray}
where $H^B_{\mu\nu}$    is given by \eqref{eq5.4}.
 
We note that, apart from the last term, the vertices which involve  the sources and the fields
are the same as those in the original Lagrangian \eqref{eq2.2}. This is a consequence  of the relation\eqref{eq3.13} 
which holds to all orders in Landau gauge, due to  the transversality of the gluon propagator.
 
Renormalization is performed by expanding the coefficients in \eqref{eq5.7} in powers of $\hbar$. To $n$-loop
order, one may adjust the $\hbar^n$ counter-terms so as to cancel the corresponding divergent contributions.
In particular, the last counter-term in \eqref{eq5.7} would then cancel the divergences arising from the Feynman graphs,
which have the structure \eqref{eq2.12}  ( compare with \eqref{eq4.8}).

\newpage

\section{Conclusion}
We have examined, in the Landau gauge, the renormalization of an alternative form of  the first order YM theory,
where the tree matrix-propagator of the $A_\mu$  and $H_{\mu\nu}$ fields is diagonal (for a simple scalar model of the diagonal formulation
\eqref{eq1.4} of the YM theory, see Appendix D).
The renormalization has been studied in the context of the BRST identities
and involves re-scalings of the fields and sources, as well as non-linear mixings.
We have shown that this theory is renormalisable to all orders and have given the form of the complete renormalized Lagrangian.
We point out that  the Landau gauge has been used just for the sake of simplicity. 
But the renormalizability also holds in a general covariant gauge, although the computations (see Appendix B) and the arguments are more
involved in this case. However, since the BRST identities preserve the gauge invariance of the theory, the proof of renormalizability
can generally be carried out by using a similar procedure.
We expect that the extension of such an approach to the first order form of the Einstein-Hilbert action
may be useful,    despite     
the fact that this gauge theory is 
not renormalisable in the usual sense. 

\acknowledgements
F. T. B. and J. F. would like to thank Professor J. C. Taylor for a helpful correspondence and CNPq (Brazil) for a grant.
D. G. C. M. would like to thank Roger Macleod for a helpful suggestion,
Fapesp (Brazil) for financial support (grant number 2018/01073-5) and Universidade de S\~ao Paulo for the hospitality.


\appendix

\section{}
Here we present the Feynman rules generated from the Lagrangian given by Eqs. \eqref{eq2.1} and \eqref{eq2.2}.
We use the standard procedure to obtain  the momentum space Feynman rules from $i S$, where $S$ is the action.
In general, whenever we have 
\be\label{eqA.A.1}
i S = \sum_n iS^{(n+m)},
\ee
where
\be\label{eqA.2}
i S^{(n+m)} = V_{i_1  \cdots i_n l_1\cdots l_m} A_{i_1} \cdots A_{i_n} F_{l_1} \cdots F_{l_m} , 
\ee
($i,l=\{a,[\mu,\nu,\dots],p\}$ is a collective index for color, Lorentz indices and momentum of the Bosonic fields $A$ and $F$),
then  
\be\label{eqA.3} 
\frac{\delta^n (i S^{(n+m)})}{\delta A_{j_1} \cdots \delta A_{j_n}  \delta F_{p_1} \cdots \delta F_{p_n}}  
=  \left[V_{j_1 \dots j_n  p_1 \dots p_m}  + \mbox{permutations of } (j_1\dots j_n) \right] + \mbox{permutations of } (p_1\dots p_m) ,
\ee
gives all the interactions vertices ($n+m\ge 2$) and 
\be\label{eqA.4} 
-\left(\frac{\delta^2 (i S^{(2)})}{\delta \phi_i\delta \phi_j}\right)^{-1}; \; \phi = A \mbox{ or }  F
\ee
yields the propagators. Of course this can be easily modified when the action contains anti-commting fields.


Proceeding in this way, we obtain from Eq. \eqref{eq2.1} the following propagators for the fields $H^a_{\mu\nu}$, $A^a_\mu$ and the ghost field $\eta^a$ 
\begin{eqnarray}\label{eqA.5}
{}^a{}^{,\mu\nu}\includegraphics[scale=0.4,trim=0.45 0.0cm 0 0]{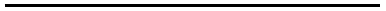}{}^b{}^{,\alpha\beta}
= 2 i \delta^{ab} \eta_{\alpha\mu} \eta_{\bt\nu},
\end{eqnarray}
\begin{eqnarray} \label{eqA.6}
{}^a{}^{,\mu}\includegraphics[scale=0.4,trim=0.45 0.0cm 0 0]{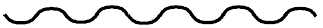}{}^b{}^{,\nu}
= -i\frac{\delta^{ab}}{p^2+i\epsilon} \left[ {\eta_{\mu\nu}} - \left(1-\xi\right)\frac{p_\mu p_\nu}{p^2+i\epsilon}\right],
\end{eqnarray}
and
\begin{eqnarray}\label{eqA.7}
{}^a{}^{}\includegraphics[scale=0.4,trim=0.45 0.0cm 0 0]{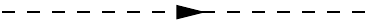}{}^b{}^{}
= \frac{i \delta^{ab}}{p^2+i\epsilon}. 
\end{eqnarray}

From the interaction terms in Eq. \eqref{eq2.2} the Feynman rules for the tree vertices
$HAA$, $AAA$ and $\bar\eta A \eta$ are respectively given by
\begin{eqnarray} \label{eqA.8}
{{{}^c{}^{,\mu\nu} }  \atop {{{}_a{}_{,\alpha} \;}\includegraphics[scale=0.6,trim=0.45 0.2cm 0 0]{treeFAA}{\;{}_b{}_{,\beta} }}}
=- \frac{ig  f^{abc}}{2} \left(\eta^{\mu\alpha} \eta^{\nu\beta} - \eta^{\mu\beta} \eta^{\nu\alpha} \right)
\end{eqnarray}
\begin{eqnarray} \label{eqA.9}
{{{}^c{}^{,\alpha,} {}^{\,p_3}}  \atop {{{}_a{}_{,\mu,}{}_{\,p_1} \;}
\includegraphics[scale=0.6,trim=0.45 0.2cm 0 0]{treeAAA}{\;{}_b{}_{,\nu}} {}_{\,p_2}}  }
= g  f^{abc} \left[\eta^{\mu\nu} (p_1-p_2)^{\alpha}  + \eta^{\alpha\nu} (p_2-p_3)^{\mu}  + \eta^{\mu\alpha} (p_3-p_1)^{\nu}  \right]
\end{eqnarray}
\begin{eqnarray} \label{eqA.10}
{{{}^b{}^{,\mu,} {}^{\,p_2}}  \atop {{{}_c{}_{,}{}_{\,p_3} \;}
\includegraphics[scale=0.6,trim=0.45 0.2cm 0 0]{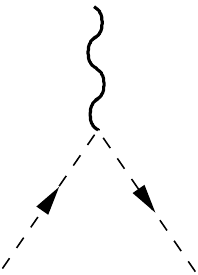}{\;{}_a{}_{,}} {}_{\,p_1}}  }
= g f^{abc} p_1^{\mu} ,
\end{eqnarray}
where Dirac delta functions enforcing momentum conservation is to be understood. 
For completness, we also present the quartic vertex generated from the Lagrangian  in Eq. \eqref{eq1.2}. 
Using the same conventions as in the previous expressions, we obtain
\begin{eqnarray}\label{eqA.11}
{                    
{{}^d{}^{,\rho} \;\;\;\;\;\;\;\;\;\;\;\;\;\;\;\,
{}^c{}^{,\nu} 
 }  \atop {{{}_a{}_{,\lambda} \;}\includegraphics[scale=0.6,trim=0.45 0.2cm 0 0.5cm]{treeAAAA}{\;{}_b{}_{,\mu} }}}
= -i g^2 \left[f^{ade}  f^{bce}   (\eta_{\lambda\mu}\eta_{\nu\rho}-\eta_{\lambda\nu}\eta_{\mu\rho})\right]
 + (c,\nu)\leftrightarrow(d,\rho)   + (b,\mu)\leftrightarrow(d,\rho).
\end{eqnarray}
Of course this quartic vertex is not present in the tree level Lagrangian of the first order formalism. 



\section{}
Here we present the results for the UV divergent part of the one-loop 1PI Green functions, up to the four gluon vertex. 
We use dimensional regularization with spacetime dimension $d=4-2 \epsilon$ and the Feynman rules presented in the 
the Appendix A. 
We express the one-loop 1PI Green functions in terms of the factor $d_N =  {g^2 N}/{16\pi^2\epsilon}$ introduced in Eq. \eqref{eq2.4}.

\subsection{One-loop two-point functions}


A straightfowrd one-loop calculation yields the following expressions for the UV contributions to the self-energies of the $H$ and $A$ fields,
as well as the mixed $\langle H A \rangle$ one:
\begin{eqnarray}\label{eqB.1}
\Pi_{\mu\nu;\;\alpha\beta}^{ab} & = &  \frac 1 2 \left({}^{a\,\mu\nu}\includegraphics[scale=0.4,trim=0.45 1.5cm 0 0]{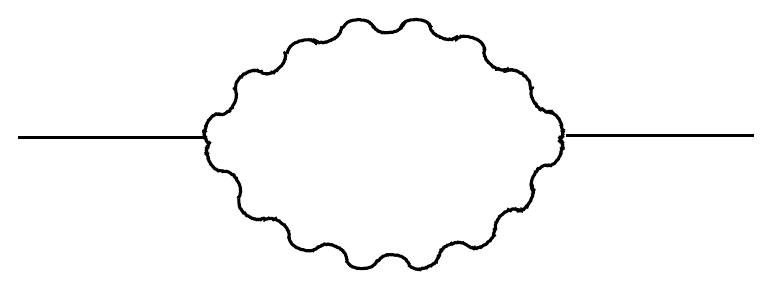}{}^{b\,\alpha\beta}\right) \nonumber \\  \\ \nonumber
&=& d_N \, 
\frac{1  + \xi}{2}
\left[\frac{i \delta^{ab}}{4}{(\eta_{\mu\alpha}\eta_{\nu\beta} - \eta_{\mu\beta}\eta_{\nu\alpha} )}\right] , 
\end{eqnarray}

\begin{eqnarray}\label{eqB.2}
\Pi_{\mu\nu}^{ab}(k) & = & \frac 1 2 \left(
{}^{a\,\mu\,k} \includegraphics[scale=0.4,trim=0 1.5cm 0 0]{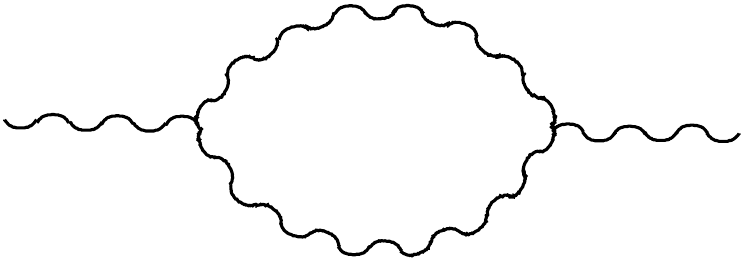} {}^{b\,\nu,\, -k}\right)\; +\; 
{}^{a\,\mu\,k} \includegraphics[scale=0.4,trim=0 1.5cm 0 0]{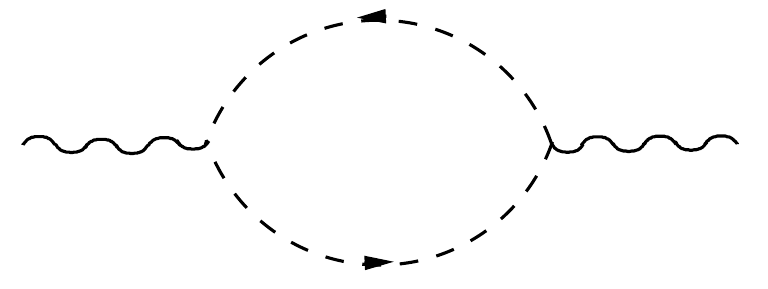} {}^{b\,\nu,\,-k}   \nonumber \\ \nonumber \\ \\ \nonumber 
& = & d_N \, 
\left(-\frac{13}{6} +\frac \xi 2\right) 
\left[ i  \delta^{ab}{(k_\mu k_\nu - k^2\eta_{\mu\nu} )} \right],
\end{eqnarray}
and
\begin{eqnarray}\label{eqB.3}
 \Pi_{\gamma;\, \mu\nu}^{ab}(k) &=&  {}^{a\,\gamma\, k}\includegraphics[scale=0.4,trim=0.45 1.5cm 0 0]{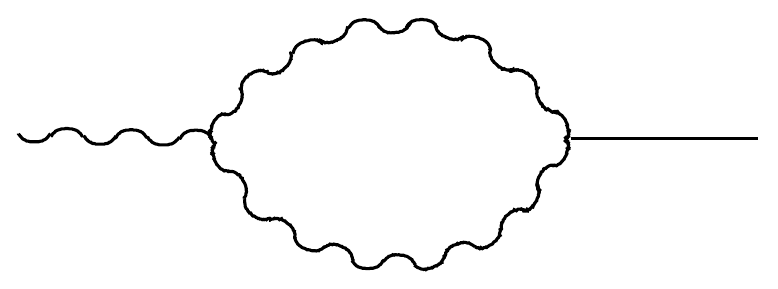}{}^{b\,\mu\nu} \nonumber \\  \\ \nonumber
 & = & d_N 
 \frac{5  + \xi}{8}\left[
 \delta^{ab} {(k_\nu \eta_{\gamma\mu} - k_\mu\eta_{\gamma\nu} )}\right],
\end{eqnarray}
where we have made explicit the symmetry factors and have employed the relation
\be
f^{a e g} f^{b g e}  = -N \delta^{ab}.
\ee

Since the diagram with one internal $H$ line vanishes in dimensional regularization (the propagator of the $F$ field is momentum independent), 
\begin{eqnarray}\label{eqB.4}
\includegraphics[scale=0.4,trim=0.5cm 1.7cm 0 0]{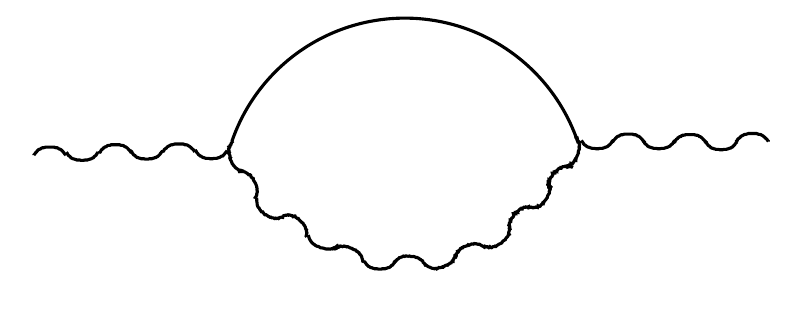}   =   0 ,  \\ \nonumber
\end{eqnarray}
we conclude that the full one-loop contribution to the 
gauge field self-energy is given by Eq. \eqref{eqB.2}, being identical to the well known result in the 
second order formalism.


The tensors inside the square brackets in Eq. \eqref{eqB.1} and \eqref{eqB.2} are identical to the momentum space representation of 
the quadratic terms in $(i S)$. The square bracket in Eq. \eqref{eqB.3} is the same as the momentum space representation of the 
mixed bi-linear term in the Lagrangian in Eq. \eqref{eq1.1}. 


\subsection{One-loop vertex corrections}

The results of the compuation of the individual graphs which contribute to the UV behaviour of the 1PI three-gluon vertex are the following
\begin{eqnarray}\label{eqB.5}
& \displaystyle{{{{}^c{}^{,\alpha,\,k_3} }  \atop {{{}_a{}_{,\mu,\, k_1} \;}
\includegraphics[scale=0.6,trim=0.45 0.2cm 0 0]{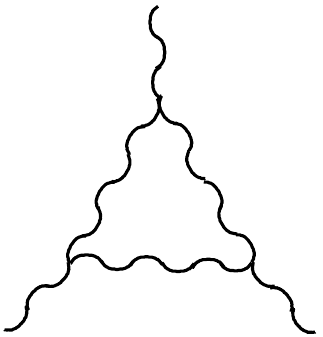}{\;{}_b{}_{,\nu,\, k_2} }}}}
= \displaystyle{d_N 
\left(   \frac{1}{2} +\frac{9 \xi}{8} \right) 
\left({{{}^c{}^{,\alpha,\,k_3} }  \atop {{{}_a{}_{,\mu,\,k_1} \;}
\includegraphics[scale=0.6,trim=0.45 0.2cm 0 0]{treeAAA}{\;{}_b{}_{,\nu,\, k_2} }}}\right) },
\\ \nonumber 
\end{eqnarray}

\begin{eqnarray}\label{eqB.6}
&\displaystyle{ {{{}^c{}^{,\alpha,\,k_3} }  \atop {{{}_a{}_{,\mu,\,k_1} \;}
\includegraphics[scale=0.6,trim=0.45 0.2cm 0 0]{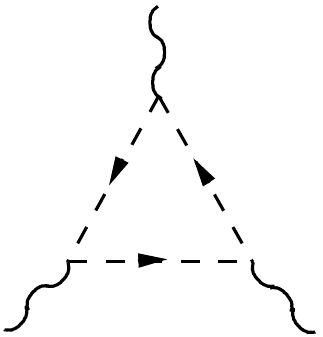}{\;{}_b{}_{,\nu,\,k_2} }}}}
+
\displaystyle{
{{{}^c{}^{,\alpha} }  \atop {{{}_a{}_{,\mu} \;}\includegraphics[scale=0.6,trim=0.45 0.2cm 0 0]{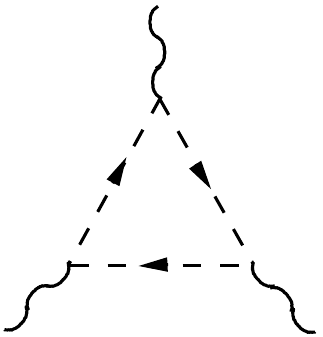}{\;{}_b{}_{,\nu} }}}}
= -\displaystyle{  \frac{d_N }{24}   
\left({{{}^c{}^{,\alpha,\,k_3} }  \atop {{{}_a{}_{,\mu,\,k_1} \;}
\includegraphics[scale=0.6,trim=0.45 0.2cm 0 0]{treeAAA}{\;{}_b{}_{,\nu,\,k_2} }}}\right) },
\\ \nonumber 
\end{eqnarray}
and
\begin{eqnarray}\label{eqB.7}
\left({{{}^c{}^{,\alpha,\,k_3} }  \atop {{{}_a{}_{,\mu,\,k_1} \;}
\includegraphics[scale=0.6,trim=0.45 0.2cm 0 0]{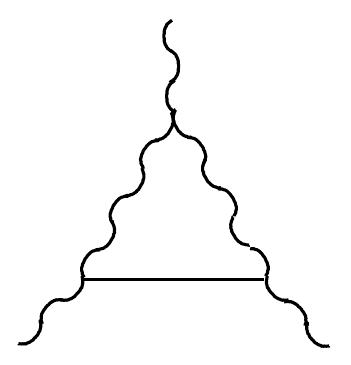}{\;{}_b{}_{,\nu,\,k_2} }}}
 \mbox{  \bf +}  \mbox{   2 permutations}\right) 
= \displaystyle{- d_N  \left(\frac{ 5 }{8} + \frac \xi 8 \right)   } 
\left({{{}^c{}^{,\alpha,\,k_3} }  \atop {{{}_a{}_{,\mu,\,k_1} \;}
\includegraphics[scale=0.6,trim=0.45 0.2cm 0 0]{treeAAA}{\;{}_b{}_{,\nu,\,k_2} }}}\right) .
\\ \nonumber
\end{eqnarray}

Adding Eqs. \eqref{eqB.5}, \eqref{eqB.6} and \eqref{eqB.7}, yields
\begin{eqnarray}\label{eqB.8}
  \displaystyle{
\left({{{}^c{}^{,\alpha,\,k_3} }  \atop {{{}_a{}_{,\mu,\,k_1} \;}
\includegraphics[scale=0.6,trim=0.45 0.2cm 0 0]{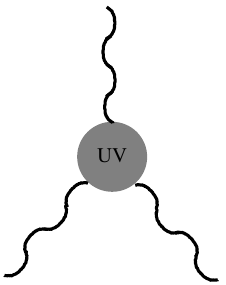}{\;{}_b{}_{,\nu,\,k_2} }}}\right) 
  }   = 
\displaystyle{d_N  \left(-\frac{ 1 }{6} + \xi  \right)   } 
\left({{{}^c{}^{,\alpha,\,k_3} }  \atop {{{}_a{}_{,\mu,\,k_1} \;}
\includegraphics[scale=0.6,trim=0.45 0.2cm 0 0]{treeAAA}{\;{}_b{}_{,\nu,\,k_2} }}}\right) 
\nonumber \\
\end{eqnarray}

The results for the individual UV contributions to the $\langle HAA\rangle$ vertex are
\begin{eqnarray}\label{eqB.9}
& 
\displaystyle{{{{}^c{}^{,\mu\nu} }  \atop {{{}_a{}_{,\alpha} \;}\includegraphics[scale=0.6,trim=0.45 0.2cm 0 0]{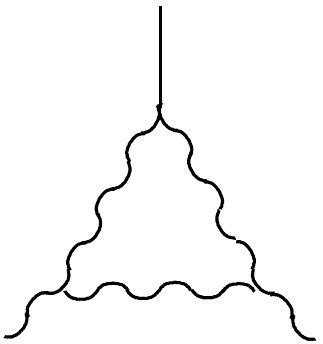}{\;{}_b{}_{,\beta} }}}}
& = d_N \displaystyle{\frac{3\xi}{4}
\left({{{}^c{}^{,\mu\nu} }  \atop {{{}_a{}_{,\alpha} \;}\includegraphics[scale=0.6,trim=0.45 0.2cm 0 0]{treeFAA}{\;{}_b{}_{,\beta} }}}\right)}
\\ \nonumber 
\end{eqnarray}
and
\begin{eqnarray}\label{eqB.10}
& \displaystyle{
{{{}^c{}^{,\mu\nu} }  \atop {{{}_a{}_{,\alpha} \;}\includegraphics[scale=0.6,trim=0.45 0.2cm 0 0]{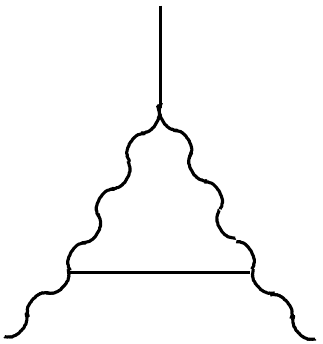}{\;{}_b{}_{,\beta} }}}}
& = \displaystyle{ d_N \left(-\frac{ 1}{4} - \frac{\xi}{4} \right) 
\left({{{}^c{}^{,\mu\nu} }  \atop {{{}_a{}_{,\alpha} \;}\includegraphics[scale=0.6,trim=0.45 0.2cm 0 0]{treeFAA}{\;{}_b{}_{,\beta} }}}\right)}
\\ \nonumber
\end{eqnarray}\label{eqB.9pluseqB.10}
Adding Eqs. \eqref{eqB.9} and \eqref{eqB.10},  we obtain the follwing result
\be
\displaystyle{
{{{}^c{}^{,\mu\nu} }  \atop {{{}_a{}_{,\alpha} \;}\includegraphics[scale=0.6,trim=0.45 0.0cm 0 0]{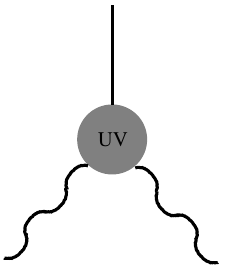}{\;{}_b{}_{,\beta} }}}}
=\displaystyle{ d_N \left(-\frac{ 1}{4} + \frac{\xi}{2} \right) 
\left({{{}^c{}^{,\mu\nu} }  \atop {{{}_a{}_{,\alpha} \;}\includegraphics[scale=0.6,trim=0.45 0.0cm 0 0]{treeFAA}{\;{}_b{}_{,\beta} }}}\right)}.
\ee
In all these expressions we have made use of
\be
f^{a e g} f^{b g h} f^{c h e} = \frac{N}{2}  f^{abc}.
\ee

Finally, let us consider the UV behaviour of the four gluon vertex. 
In order to remark some fine details involving cancellations, let us consider the contributions of individual diagrams.
We have performed the calculations using the SU(2) structure constants,
which is suficiently general in order to obtain the UV structures. Then, in order to convert the to SU(N), we multiply the
final result by the factor $N/2$. 

The simplest UV contribution comes from the ghost loop diagrams, which gives the following result 
\begin{eqnarray}\label{eqB.11}
& \nonumber \\ &
\displaystyle{
{                    
{{}^d{}^{,\rho} \;\;\;\;\;\;\;\;\;\;\;\;\;\;\;\;\;\;\;
{}^c{}^{,\nu} 
 }  \atop {{{}_a{}_{,\lambda} \;}\includegraphics[scale=0.6,trim=0.45 -0.3cm 0 0]{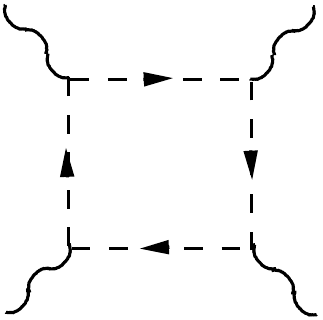}{\;{}_b{}_{,\mu} }}}
+
{                    
{{}^d{}^{,\rho} \;\;\;\;\;\;\;\;\;\;\;\;\;\;\;\;\;\;\;
{}^c{}^{,\nu} 
 }  \atop {{{}_a{}_{,\lambda} \;}\includegraphics[scale=0.6,trim=0.45 -0.3cm 0 0]{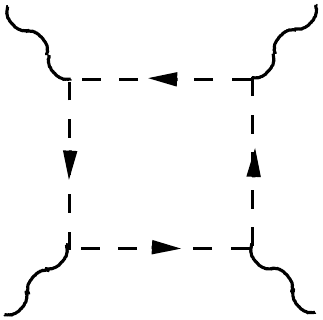}{\;{}_b{}_{,\mu} }}}}
+\mbox{ 2 permutations}    \nonumber \\  & \\ \nonumber & 
= \displaystyle{- d_2 \, 
\frac{i g^2}{12} \,  \left(\delta^{ad} \delta^{bc}+\delta^{ac} \delta^{bd}+\delta^{ab} \delta^{cd}\right)
               \left(\eta_{\lambda \rho }\eta_{\mu \nu }+\eta_{\lambda \nu } \eta_{\mu \rho }+\eta_{\lambda \mu } \eta_{\nu \rho}\right)}.
\end{eqnarray}

The graphs with two and one  momentum independent internal $F$ propagators yields respectively the following results
\begin{eqnarray}\label{eqB.12}
& \nonumber \\ &
\displaystyle{
{                    
{{}^d{}^{,\rho} \;\;\;\;\;\;\;\;\;\;\;\;\;\;\;\;\;\;\
{}^c{}^{,\nu} }  \atop {{{}_a{}_{,\lambda} \;}\includegraphics[scale=0.6,trim=0.45 -0.3cm 0 0]{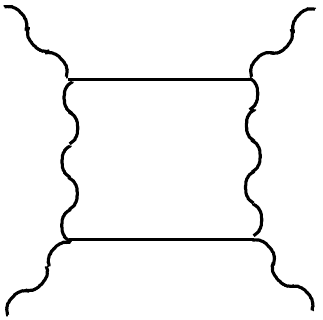}{\;{}_b{}_{,\mu} }}}}
+ \mbox{ 5 permutations} \nonumber \\  &  \nonumber \\ & 
= \displaystyle{ d_2 \, \frac{i g^2}{12} } 
\left\{
\delta^{ab} \delta^{cd} \left[\left(4 \xi ^2+\xi +13\right) \left(\eta_{\lambda \rho } \eta_{\mu \nu
   }+\eta_{\lambda \nu } \eta_{\mu \rho }\right)+(7 \xi^2  + 4\xi +25) \eta_{\lambda \mu } \eta_{\nu   \rho }\right]
 \right. \nonumber \\   
& \;\;\;\;\;\;\;\;\;\;\;\; +
 \delta^{ac} \delta^{bd} \left[\left(4 \xi ^2+\xi +13\right) \left(\eta_{\lambda \rho } \eta_{\mu \nu }+\eta_{\lambda \mu } \eta_{\nu \rho } \right)
   +\left(7   \xi ^2+4 \xi +25\right) \eta_{\lambda \nu } \eta_{\mu \rho } \right]
 \nonumber \\ 
& \;\;\;\;\;\;\;\;\;\;\;\; + \left.
 \delta^{ad} \delta^{bc} \left[\left(4 \xi ^2+\xi +13\right) \left(\eta_{\lambda \nu } \eta_{\mu \rho }+\eta_{\lambda \mu }\eta_{\nu \rho }\right) +
   \left(7 \xi ^2+4 \xi +25\right) \eta_{\lambda \rho } \eta_{\mu \nu}
   \right] \right\}
\end{eqnarray}
and
\begin{eqnarray}\label{eqB.13}
& \nonumber \\ &
\displaystyle{                    
{{}^d{}^{,\rho} \;\;\;\;\;\;\;\;\;\;\;\;\;\;\;\;\;\;\;\;\;\;\;
{}^c{}^{,\nu} }  \atop {{{}_a{}_{,\lambda} \;}\includegraphics[scale=0.6,trim=0.45 -0.3cm 0 0]{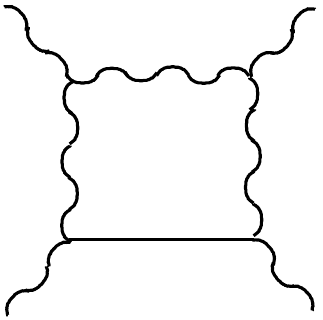}{\;{}_b{}_{,\mu} }}}
+ \mbox{ 11 permutations} \nonumber \\  &  \nonumber \\ & 
= -\displaystyle{d_2 \, \frac{i g^2}{12} } 
\left\{
\delta^{ab} \delta^{cd}
\left[\left(8 \xi ^2 - \xi +32\right) \left(\eta_{\lambda \rho } \eta_{\mu \nu}+\eta_{\lambda \nu } \eta_{\mu \rho }\right)
  +2 (7 \xi^2  + 16\xi +28) \eta_{\lambda \mu } \eta_{\nu   \rho }\right]
 \right. \nonumber \\   
& \;\;\;\;\;\;\;\;\;\;\;\; +
 \delta^{ac} \delta^{bd} \left[\left(8 \xi ^2 - \xi +32\right) \left(\eta_{\lambda \rho } \eta_{\mu \nu}+\eta_{\lambda \mu } \eta_{\nu \rho }\right)
  +2 (7 \xi^2  + 16\xi +28) \eta_{\lambda \nu } \eta_{\mu   \rho }\right]
 \nonumber \\ 
& \;\;\;\;\;\;\;\;\;\;\;\; + \left.
 \delta^{ad} \delta^{bc} 
\left[\left(8 \xi ^2 - \xi +32\right) \left(\eta_{\lambda \mu } \eta_{\rho \nu}+\eta_{\lambda \nu } \eta_{\mu \rho }\right)
  +2 (7 \xi^2  + 16\xi +28) \eta_{\lambda \rho } \eta_{\mu   \nu }\right]
 \right\} .
\end{eqnarray}
Finally the contribution from the diagram with four internal gluons lines is given by
\begin{eqnarray}\label{eqB.14}
& \nonumber \\ &
\displaystyle{                    
{{}^d{}^{,\rho} \;\;\;\;\;\;\;\;\;\;\;\;\;\;\;\;\;\;\;\;\;\;
{}^c{}^{,\nu} }  \atop {{{}_a{}_{,\lambda} \;}\includegraphics[scale=0.6,trim=0.45 -0.3cm 0 0]{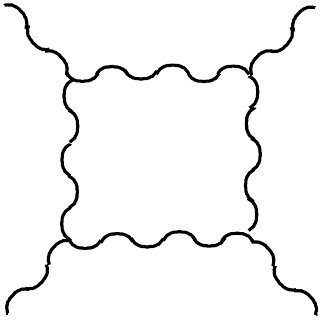}{\;{}_b{}_{,\mu} }}}
+ \mbox{ 2 permutations} \nonumber \\  &  \nonumber \\ & 
= \displaystyle{d_2 \, \frac{i g^2}{12} } 
\left\{
\delta^{ab} \delta^{cd}
\left[\left(4 \xi ^2 + 4 \xi + 24\right) \left(\eta_{\lambda \rho } \eta_{\mu \nu}+\eta_{\lambda \nu } \eta_{\mu \rho }\right)
  +(7 \xi^2  + 16\xi + 24) \eta_{\lambda \mu } \eta_{\nu   \rho }\right]
 \right. \nonumber \\   
& \;\;\;\;\;\;\;\;\;\;\;\; +
 \delta^{ac} \delta^{bd} \left[\left(4 \xi ^2 + 4 \xi + 24\right) \left(\eta_{\lambda \rho } \eta_{\mu \nu}+\eta_{\lambda \mu } \eta_{\nu \rho }\right)
  + (7 \xi^2  + 16\xi + 24) \eta_{\lambda \nu } \eta_{\mu   \rho }\right]
 \nonumber \\ 
& \;\;\;\;\;\;\;\;\;\;\;\; + \left.
 \delta^{ad} \delta^{bc} 
\left[\left(4 \xi ^2 +  4 \xi + 24\right) \left(\eta_{\lambda \mu } \eta_{\rho \nu}+\eta_{\lambda \nu } \eta_{\mu \rho }\right)
  + (7 \xi^2  + 16\xi + 24) \eta_{\lambda \rho } \eta_{\mu   \nu }\right]
 \right\}.
\end{eqnarray}
Adding Eqs. \eqref{eqB.11}, \eqref{eqB.12}, \eqref{eqB.13} and \eqref{eqB.14}, one can see that the $\xi^2$ terms cancel and we are
left with a result which can be expressed in terms of the tree level four gluon vertex, given by Eq. \eqref{eqA.11}, as follows
\begin{eqnarray}\label{eqB.15}
  &
\displaystyle{                    
{{}^d{}^{,\rho} \;\;\;\;\;\;\;\;\;\;\;\;\;\;\;
{}^c{}^{,\nu} 
 }  \atop {{{}_a{}_{,\lambda} \;}\includegraphics[scale=0.6,trim=0.45 -0.3cm 0 0]{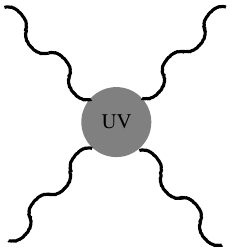}{\;{}_b{}_{,\mu} }}}
& =
 d_N \; \displaystyle{  \left(\frac 1 3 + \frac{\xi}{2} \right)} \;
\left(
\displaystyle{                    
{{}^d{}^{,\rho} \;\;\;\;\;\;\;\;\;\;\;\;\;\;\;\,
{}^c{}^{,\nu} 
 }  \atop {{{}_a{}_{,\lambda} \;}\includegraphics[scale=0.6,trim=0.45 0.2cm 0 0.5cm]{treeAAAA}{\;{}_b{}_{,\mu} }}}
\right),
%
\\ \nonumber 
\end{eqnarray}
where we have multiplied the final result by $d_N/d_2$. 

\section{}
Here we show that the transformations \eqref{eq5.3} -\eqref{eq5.6} preserve, to all orders, the BRST  condition
\be\label{eqC.1}
\Gamma^R\star\Gamma^R = \int d^4 x\left[
\frac{\delta\Gamma^R}{\delta H_{\mu\nu}} \cdot \frac{\delta\Gamma^R}{\delta \Omega^{\mu\nu}} +
\frac{\delta\Gamma^R}{\delta A_{\mu}} \cdot \frac{\delta\Gamma^R}{\delta U^{\mu}} +
\frac{\delta\Gamma^R}{\delta \eta} \cdot \frac{\delta\Gamma^R}{\delta V}
\right] = 0. 
\ee
Evaluating the three terms in \eqref{eqC.1} , one gets in the Landau gauge
\begin{subequations}
\begin{eqnarray}
\frac{\delta\Gamma^R}{\delta H_{\mu\nu}} \cdot \frac{\delta\Gamma^R}{\delta \Omega^{\mu\nu}} 
& = & \left( Z_A Z_\eta \right)^{1/2} 
\frac{\delta\Gamma^R}{\delta H^B_{\mu\nu}} \cdot \frac{\delta\Gamma^R}{\delta \Omega^{B \mu\nu}} \nonumber \\
&+& 2 Z_\eta^{1/2} \frac{\delta\Gamma^R}{\delta H^B_{\mu\nu}}  \cdot \left[ 
Z_{AH} \partial_\nu\frac{\delta\Gamma^R}{\delta U^{B\,\mu}} - g Z_{AAH} \frac{\delta\Gamma^R}{\delta U^{B\,\mu}} \wedge A_\nu 
\right] \label{eqC.2} \\
\frac{\delta\Gamma^R}{\delta A_{\mu}} \cdot \frac{\delta\Gamma^R}{\delta U^{\mu}} 
& = & \left( Z_A Z_\eta \right)^{1/2}  
\frac{\delta\Gamma^R}{\delta A^B_{\mu}} \cdot \frac{\delta\Gamma^R}{\delta U^{B \mu}} \nonumber \\ 
&+ & 2 Z_\eta^{1/2} \frac{\delta\Gamma^R}{\delta U^{B\,\mu}} \cdot \left[ 
Z_{AH} \partial_\nu\frac{\delta\Gamma^R}{\delta H^{B}_{\mu\nu}} - g Z_{AAH} 
\frac{\delta\Gamma^R}{\delta H^{B}_{\mu\nu}} \wedge A_\nu \right]\label{eqC.3} \\
\frac{\delta\Gamma^R}{\delta\eta} \cdot \frac{\delta\Gamma^R}{\delta V} 
& = & \left( Z_A Z_\eta\right)^{1/2}  \frac{\delta\Gamma^R}{\delta\eta^B} \cdot \frac{\delta\Gamma^R}{\delta V^B}   \label{eqC.4}
\end{eqnarray}
\end{subequations}
Adding these equations, we see that the last two terms in \eqref{eqC.2} and \eqref{eqC.3} cancel out.
Hence,the sum gives
\begin{eqnarray}\label{eqC.5}
\frac{\delta\Gamma^R}{\delta H_{\mu\nu}} \cdot \frac{\delta\Gamma^R}{\delta \Omega^{\mu\nu}} +
\frac{\delta\Gamma^R}{\delta A_{\mu}} \cdot \frac{\delta\Gamma^R}{\delta U^{\mu}} +
\frac{\delta\Gamma^R}{\delta \eta} \cdot \frac{\delta\Gamma^R}{\delta V} 
= \nonumber \\ \left(Z_A Z_\eta\right)^{1/2}\left[
\frac{\delta\Gamma^R}{\delta H^B_{\mu\nu}} \cdot \frac{\delta\Gamma^R}{\delta \Omega^{B\mu\nu}} +
\frac{\delta\Gamma^R}{\delta A^B_{\mu}} \cdot \frac{\delta\Gamma^R}{\delta U^{B\mu}} +
\frac{\delta\Gamma^R}{\delta \eta^B} \cdot \frac{\delta\Gamma^R}{\delta V^B} 
\right] = 0
\end{eqnarray}
where we used the fact that (see \eqref{eq5.2}) 
$\Gamma^R = \Gamma_0(g^B;A^B,\eta^B,\bar\eta^B, H^B; U^B, V^B, \Omega^B)$
obeys the bare BRST equation. Using this result, it follows that  $\Gamma^R$       satisfies
the BRST equation \eqref{eqC.1}.

\section{}
Here we present a simple scalar model for the diagonal formulation \eqref{eq1.4} of the YM theory.
Let us consider the massless $\lambda \phi^4$ model described by the Lagrangian
\be\label{eqD.1}
{\cal L} = \frac 1 2 (\partial^\mu\phi)(\partial_\mu\phi) - \lambda \phi^4.
\ee
We add to this a term involving an auxiliary $B$ field, which leads to 
\be\label{eqD.2}
{\cal L}^{1} = \frac 1 2 (\partial^\mu\phi)(\partial_\mu\phi) - \lambda \phi^4 + \lambda B^2.
\ee
Using the equation of motion for the $B$-field, we get at the classical level that $B=0$, so that \eqref{eqD.2} is equivalent
to \eqref{eqD.1}. If we set $B=\sqrt{1/(2\lambda)} H + \phi^2$, we obtain
\be\label{eqD.3}
{\cal L}^{2} = \frac 1 2 H^2 + \frac 1 2 (\partial^\mu\phi)(\partial_\mu\phi) + \sqrt{2\lambda} H \phi^2,
\ee
which involves only a cubic coupling $H\phi^2$ and is rather similar to the diagonal formulation \eqref{eq1.4} of the YM theory.
We note that if we employ \eqref{eqD.3} in the path integral formulation of the generating functional, the integration over $H$
yields the same Green's functions as the ones generated from the original Lagrangian \eqref{eqD.1}.



\end{document}